\documentclass[usenatbib,usegraphicx]{mn2e}
\usepackage{times}

\title{Dark matter annihilation in the halo of the Milky Way}
\author[F. Stoehr et al.]
{Felix Stoehr$^1$, Simon D.~M. White$^1$, Volker Springel$^1$, \newauthor Giuseppe Tormen$^2$ and Naoki Yoshida$^3$\\
$^1$Max-Planck-Institut f\"{u}r Astrophysik, Karl-Schwarzschild-Str. 1, 85748 Garching 
bei M\"unchen, Germany\\
$^2$Dipartimento di Astronomia, Universita di Padova, vicolo 
dell'Osservatorio 5, 35122 Padova, Italy\\
$^3$National Astronomical Observatory Japan, Mitaka, Tokyo 181-8588, Japan\\
Email: felix@mpa-garching.mpg.de, swhite@mpa-garching.mpg.de, 
volker@mpa-garching.mpg.de, tormen@pd.astro.it, naoki@th.nao.ac.jp}
\date{Accepted 2003 July 29. Received 2003 July 7}
\newcommand{\prd}{Physical Review}
\newcommand{\mnras}{MNRAS}
\newcommand{\nat}{Nature}
\newcommand{\apj}{ApJ}
\newcommand{\aj}{AJ}

\newcommand{\apjl}{ApJL}

\begin{document}
\maketitle

\begin{abstract}
If the dark matter in the Universe is made of weakly self-interacting
particles, they may self-annihilate and emit $\gamma$-rays. We use
high resolution numerical simulations to estimate directly the
annihilation flux from the central regions of the Milky Way and from
dark matter substructures in its halo. Although such estimates remain
uncertain because of their strong dependence on the structure of the
densest regions, our numerical experiments suggest that less direct
calculations have overestimated the emission both from the centre and
from halo substructure. We estimate a maximal enhancement of at most a
factor of a few with respect to a smooth spherical halo of standard 
Navarro-Frenk-White (NFW) structure. We discuss detection strategies 
for the next generation of $\gamma$-ray detectors and find that the 
annihilation flux may be detectable, regardless of uncertainties 
about the densest regions, for the annihilation cross-sections predicted 
by currently popular elementary particle models for the dark matter.

\end{abstract}


\section{Introduction}
The nature of the dark matter (DM) in the Universe is one of the most
prominent unsolved questions in cosmology. Among the best motivated
candidates for DM is a weakly interacting massive particle (WIMP) in
the mass range $10$ to $10^4$ GeV~$c^{-2}$. In minimal supersymmetric
extensions of the standard model of particle physics, a stable,
neutral particle with these properties (usually called a neutralino)
arises naturally as the lightest supersymmetric particle.

In recent years, a growing effort has been dedicated to detecting
WIMPs directly through the energy deposited by elastic WIMP-nucleon
scattering in massive, cryogenically cooled bolometers. A different
detection strategy is possible if WIMPS are Majorana particles. In
this case, pair-annihilations can occur, producing high-energy
neutrinos, positrons, antiprotons and $\gamma$-rays. The resulting
$\gamma$-ray fluxes might be detectable with current- or
next-generation telescopes. So far neither technique has detected a DM
particle.

Both ground-based air-shower-\v{C}erenkov telescopes (ACT) and
space-borne telescopes might be able to detect annihilation
$\gamma$-rays. Such observations can be used to constrain WIMP
parameters -- in particular, the self-annihilation cross-section and the
particle mass. Predictions of the expected flux require not only these
parameters but also a detailed knowledge of the structure of regions
of high DM density, i.e. of DM haloes. This is because the
annihilation flux (in photons cm$^{-2}$ s$^{-1}$) may be written as:
\begin{equation}
\label{equation:flux}
   F = \frac{N_{\gamma}\left<\sigma
     v\right>}{2 \, \, m_{DM}^2} \int_V \frac{\rho_{DM}^2({\bf x})} 
    {4\pi \, d^2({\bf x})} \, d^3x\, ,
\end{equation}
where $N_{\gamma}$ is the number of photons produced per annihilation,
$\left<\sigma v\right>$ is the averaged product of cross-section and
relative velocity, $\rho_{DM}$ is the DM density, $V$ is the halo
volume, $m_{DM}$ is the mass of the DM particle and $d$ the distance
from each point in the halo to the observer. The density squared
weighting of the integrand in this equation results in most of the
flux in dark DM haloes being produced by the small fraction of
their mass in the densest regions.

Two specific regions have been suggested as dominating the
annihilation signal from haloes. A large contribution could come from
the innermost part of the halo. For a distant spherically symmetric
system equation (1) becomes
\begin{equation}
  F =  \frac {N_{\gamma}\left<\sigma v\right>}{2 \, d^2 m_{DM}^2} 
        \int_0^{r_{200}}\rho_{DM}^2(r) r^2 dr\, ,
\end{equation}
so if the inner density profile is $\propto r^{-1.5}$ or steeper, the
emitted flux diverges at the centre. A lower cut-off must then be
specified on physical grounds, for example at the point where the
annihilation timescale for the DM becomes equal to the lifetime of the
halo. This divergent case might, perhaps, be relevant, since at least
some high resolution numerical simulations have suggested that the
inner cusps of DM haloes could be this steep \citep{Moore_et_al_99,
Calcaneo_Moore_00}. Both the data we present below and the study of
\citet{Power_et_al_03} suggest, however, that cold dark matter (CDM) haloes are
substantially less concentrated than this.

A second contribution can come from small-scale structure in the DM
distribution in the halo. It is now well established that 5 to 10 per cent of
the halo mass in (CDM) haloes is contained in
gravitationally self-bound substructures
\citep{Moore_et_al_99,Klypin_et_al_99}. If the central regions of these
subhaloes were dense enough, they could produce a substantial fraction of
the total annihilation radiation from the halo
\citep{Calcaneo_Moore_00}.

In recent years, advances in integrator software, multi-mass initial
condition techniques and computer speed have made it possible to
simulate the DM halo of the Milky Way with sufficient resolution to
see the dense regions which dominate the annihilation
signal. In the present paper we use a series of high resolution N-body
DM simulations to predict the annihilation flux from a CDM halo
similar in mass to the halo of the Milky Way.

In the next section we briefly describe the N-body simulations we have
carried out. In Section 3 we measure circular velocity profiles for
our haloes and discuss their implications for the strength of the
annihilation flux from the inner Galaxy. In Section 4 we then analyse
the flux enhancements due to bound substructures and to other density
irregularities. In Section 5, we use the annihilation cross-sections currently
considered plausible to evaluate whether halo
$\gamma$-ray production is likely to be detectable with the next
generation of telescopes. Finally, Section 6 summarises our results and
compares them to those of other workers.

\section{$N$-body simulations}
\label{section:nbodysimulations}
\begin{figure}
\centering
\includegraphics[width=84mm]{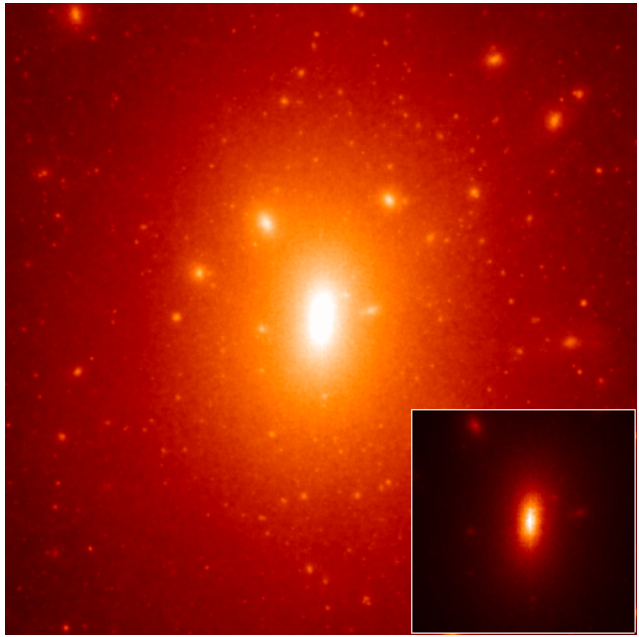}
\caption{The distribution of DM in our highest-resolution simulation
GA3n. The region displayed is a cube of side 270 kpc, i.e. 1 times
$r_{200}$. Each particle is weighted by its local density so that the
picture represents an image in annihilation radiation. The main image
has a logarithmic intensity scale, whereas the small image reproduces
the centre on a linear intensity scale.
\label{figure:dmdistribution}}
\end{figure}

\begin{figure}
\centering
\includegraphics[width=84mm]{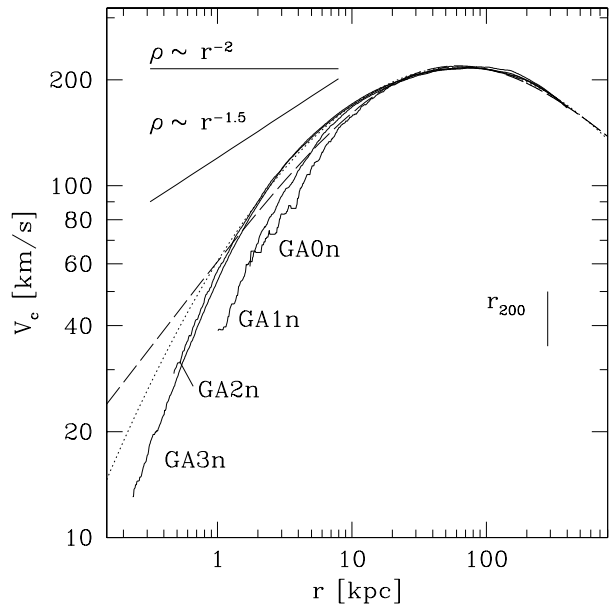}
\caption{Circular velocity curves for the simulations GA0n, GA1n, GA2n and GA3n. 
The vertical line indicates the location of the virial radius $r_{200}$. The best-fitting 
NFW profile with concentration $c_{NFW}=10$ is plotted in long dashes. A fit of the 
form proposed by SWTS with $a$=0.17 is shown in dots. At small radii the slope for 
GA3n is considerably below that corresponding to a density profile with 
$\rho \propto r^{-1.5}$.} \label{figure:rotationcurves}
\end{figure}

In this paper we use simulations of the `Milky Way' halo studied previously 
by \citeauthor{Stoehr_et_al_02} (\citeyear{Stoehr_et_al_02};
hereafter SWTS). We work with a flat $\Lambda$-dominated CDM 
universe, with matter density $\Omega_{\rm m}=0.3$,
cosmological constant $\Omega_{\Lambda}=0.7$, expansion rate $H_{0}=70
$ km~s$^{-1}$Mpc$^{-1}$, index of the initial fluctuation power
spectrum $n=1$, and present-day fluctuation amplitude
$\sigma_8=0.9$. We begin with an intermediate resolution dark matter
simulation of a `typical' region of the Universe ($N\sim 6\times
10^7$, particle mass $\sim 10^8M_\odot$) for which the techniques of
\citeauthor{Springel_et_al_01} (\citeyear{Springel_et_al_01};
hereafter SWTK) have been used to follow the formation of the galaxy
population. We identify a relatively isolated `Milky Way-like'
galaxy which had its last major merger at $z>1.5$. Then we resimulate
its halo at a series of higher resolutions, again using techniques
from \citet{Tormen_Bouchet_White_97} and SWTK and the $N$-body code
{\sc gadget}\footnote{www.mpa-garching.mpg.de/gadget/}
\citep{Springel_Yoshida_White_01}. We have rerun the simulations of
SWTS with higher force accuracy and have added an additional
simulation with even higher mass resolution. In the simulations GA0n,
GA1n, GA2n and GA3n the resimulated halo has 14~097, 128~276, 1204~411
and 10~089~396 particles, respectively within $r_{200}$, the radius
within which the enclosed mean density is 200 times the critical
value.

The simulated haloes were scaled-down in velocity, mass and radius
(but with unchanged density and time-scales) so that their circular
velocity peaks at 220~km s$^{-1}$. With this scaling, dark matter particle
masses are $1.8 \times 10^8, 1.9 \times 10^7$, $2.0 \times 10^6 $ and
$2.5 \times 10^5$ M$_{\odot}$ and Plummer equivalent softening lengths
are 1.8, 1.0, 0.48 and 0.24 kpc in GA0n, GA1n, GA2n and GA3n,
respectively. In all four $r_{200}\approx 270$~kpc. Note that since the
stars of the Milky Way contribute significantly to its measured rotation
velocity, our chosen scaling probably produces too large a mass for
the Milky Way's halo and thus also for substructures within it. We use
{\sc subfind} (SWTK) to identify self-bound substructure. A more
detailed description of the simulation set-up can be found in SWTS.
Fig.~\ref{figure:dmdistribution} shows the projected, density-weighted
DM distribution in a logarithmic representation which corresponds to
an image of its annihilation radiation. Many substructures are
visible. A representation of the halo centre with a linear intensity
scale (but on the same angular scale) is shown in the inset.

\section{Smooth halo structure}
\label{section:smoothhalo}
The most crucial parameter determining the annihilation rate in a
smooth halo is the point at which the slope of its density profile 
passes through the critical value $-1.5$. Most of the annihilation
radiation will come from this region. It is
difficult to distinguish slopes in logarithmic plots of density
against radius, so in Fig.~\ref{figure:rotationcurves} we show circular velocity profiles
\begin{equation}
V_c(r) = \left[\frac{G \ M(<r)}{r}\right]^{1/2}= \left[\frac{G}{r} \int_{V(r)}
\rho_{DM}({\bf x})\, dV \right]^{1/2} ,
\end{equation}
where $V(r)$ is the region within distance $r$ of halo centre. We plot
these curves down to a distance from the centre equal to the softening
length, and we overplot the best-fitting Navarro-Frenk-White (NFW) profile
\citep{Navarro_Frenk_White_97} in dashes. We also indicate the critical
value for the slope, $\rho(r)
\propto r^{-1.5}$, and the value for an isothermal profile $\rho(r) \propto
r^{-2}$. We find the NFW profile
\begin{equation}
\rho(r) = \, \frac{3H^2_0} {8\pi G} \,\, \frac{\delta_c}{c_{NFW}x \left(1+ c_{NFW} x \right)^2}
\end{equation}
with $x=r/r_{200}$ and $c=10$ to be a reasonably good fit to our data outside the inner core region. A better fit is a parabolic function of the form proposed by SWTS with a width parameter of  $a$= 0.17 (the dotted curve in the figure). This profile has a substantially shallower
density profile at small radii than the NFW profile. The measured circular velocity profiles of GA2n and GA3n agree very well, but a comparison with GA1n and GA0n suggests
that this apparent convergence is in part a fluke. The curves for the
two lower-resolution simulations converge to within 5 per cent of the high-resolution answer beyond about 5 times their respective softening
lengths. Assuming this to be true for GA2n and GA3n also, the inner
slope of the density profile of our simulated halo is 
established to be well below $ -1.5$. With this criterion, convergence
for GA3n is achieved just outside 1 kpc or at about 0.4 per cent of
$r_{200}$. This agrees with the expectation from the convergence study
of \citet{Power_et_al_03} who performed a large number of simulations
of several haloes using two different N-body codes as well as a wide
variety of code parameters (timestep, softening, particle number,
etc.).

\begin{figure}
\centering \includegraphics[width=84mm]{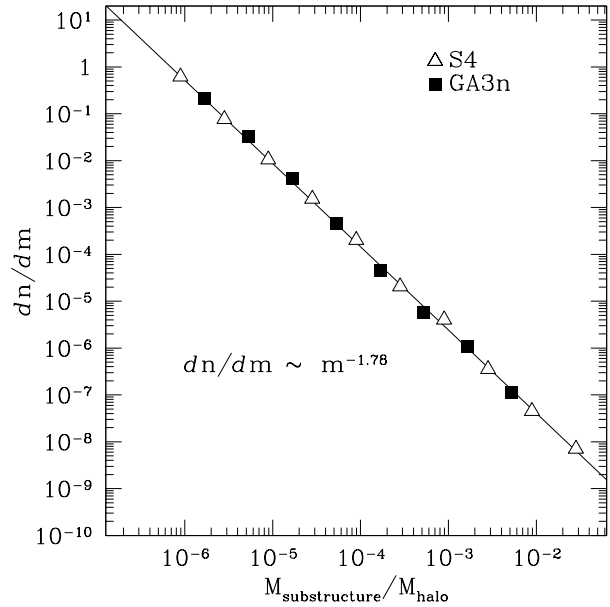}
\caption{Subhalo mass functions for the GA3n (Milky Way) and S4
(cluster) simulations. Subhaloes identified by {\sc subfind} and with
ten or more particles are included in these distributions. \label{figure:massfunctions}}
\end{figure}

The concentration parameter for our NFW fit to our halo is
$c_{NFW}=10$. Thus, $\delta_c=4.48 \times 10^4$ and the above value of
$r_{200}$ implies a scale radius $r_s=27$ kpc, and a density at the
Sun's position ($r_0$=8.0 kpc) of $\rho_0=$ $1.2 \times 10^7$
M$_\odot/$kpc$^3=$ 0.46 GeV~$c^{-2}$~cm$^{-3}$.

For our NFW fit, half of the annihilation radiation is predicted to
come from within $0.26~r_s$ which is 7 kpc. Thus the
resolution in GA3n appears easily sufficient to measure the bulk of the
emission, even though still better resolution would clearly be
desirable. Unfortunately, the numerical situation will not improve
dramatically in the next few years unless revolutionary new techniques
are discovered. As discussed at length by \citet{Power_et_al_03}, an
increase in halo particle number by (say) two orders of magnitude
would provide an increase in length resolution at halo centre by at
best a factor of 10.

Many authors have tried to determine the inner slope of dark halo
density profiles through physically based analytic arguments
\citep{Peebles_80, Hernquist_90, Syer_White_98, Nusser_Sheth_99,
Subramanian_et_al_00, Dekel_et_al_02} but despite some interesting ideas
a convincing final answer is still missing. For the purposes of this
paper the critical issue is whether the slope of the density profile
interior to the points for which it has so far been estimated
accurately from simulations (i.e. at radii below 1 kpc) remains
significantly shallower than $-1.5$. If it does, then the integral over
the smooth halo density distribution is convergent and can be
estimated reasonably accurately from high resolution simulations, for
example from GA3n which is currently the best resolved simulation of a
galaxy halo ever carried out.

\citet*{Bergstrom_et_al_98} show that if a smooth NFW
halo, similar to that of Fig.~\ref{figure:rotationcurves}, is a good description of the Milky
halo of the Way, then for some minimal super-symmetric extensions of the
Standard Model (MSSM) the $\gamma$-ray flux may just be
detectable with next generation telescopes. The flux could,
however, be significantly enhanced if the density distribution within
the halo is sufficiently clumpy \citep{Bergstrom_et_al_99,
Calcaneo_Moore_00,Taylor_Silk_03}. We now estimate whether the clumpiness of our
simulated haloes is enough to produce a large enhancement.

\section{Halo substructure}
\label{section:substructure}
In GA0n, GA1n, GA2n and GA3n the total mass in gravitationally bound
substructures, identified with {\sc subfind}, is $1.7$, $3.0$,
$5.1$ and $4.1$ per cent, respectively; the fluctuations are due to the
exclusion or inclusion of one or two massive satellites within the
radius $r_{200}$ that we consider to define the halo boundary. These
percentages are very similar to those found in the cluster simulations
of SWTK. In Fig.~\ref{figure:massfunctions} we show the abundance of
substructures as a function of mass fraction for our highest
resolution simulation GA3n, as well as for S4, the highest-resolution
simulation of SWTK.

\begin{figure*}
\centering \includegraphics[width=84mm]{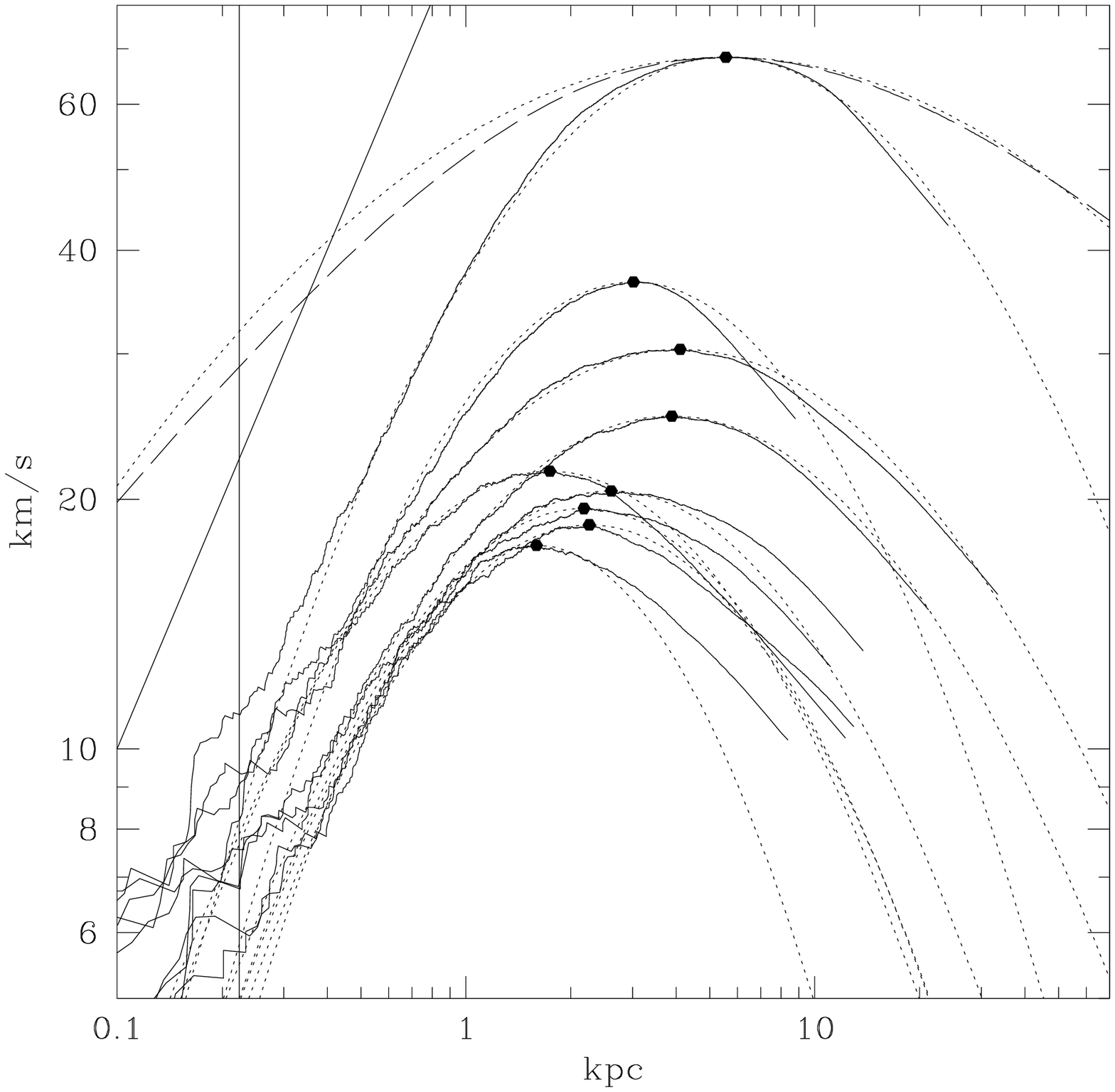}
\centering \includegraphics[width=84mm]{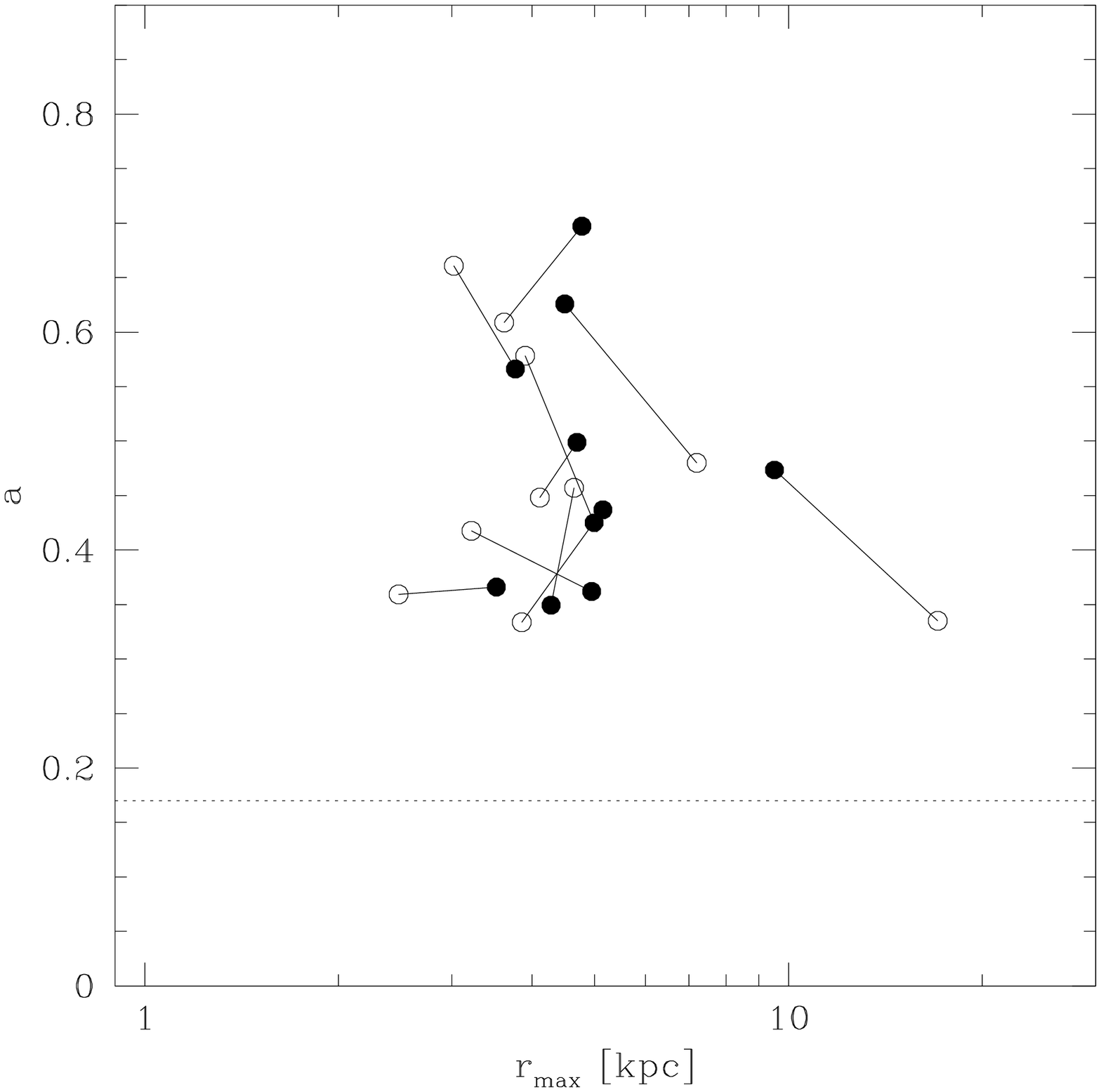}
\caption{{\it Left-hand panel:} Circular velocity curves for the GA3n subhaloes ranked 1, 5, 10,...40 in mass (solid) together with corresponding SWTS fits (dotted). For comparison an NFW profile (dashed) and an SWTS profile with $a=0.17$ (the value for the main halo) are overplotted on the most massive subhalo. The vertical solid line shows the softening length, the diagonal line the profile slope corresponding to a constant density. {\it Right-hand panel:} Values of $a$ and $r_{max}$ (the radius of maximum circular velocity) for matching subhaloes in GA2n (open) and GA3n (filled). The horizontal line is $a=0.17$, the value for the main halo. \label{figure:subhalocurves}}
\end{figure*}

These two mass functions are remarkably similar and are very close to
a power-law $dn/dm \propto m^{-1.78}$ as shown by the solid line in
the figure. They are consistent with the findings of other authors
\citep{Moore_et_al_99, Klypin_et_al_99, Metcalf_Madau_01, Font_et_al_01,
Helmi_et_al_02}, although the range of published values (1.75 to 1.9) suggests
that the close agreement between these two particular haloes is likely
to be a fluke. Some variation is undoubtedly due to the fact that
different authors use different algorithms to define substructure, but
these effects have not yet been studied in detail. Note that such
slopes are also found at low mass for the mass function of isolated
haloes in a $\Lambda$CDM Universe, suggesting that the fraction of
mass lost depends only weakly on the initial mass of an accreted DM
halo. These slopes are shallow enough to ensure that most of the mass
in substructures is contained in the few most massive objects. Thus we
do not expect the total mass in substructure to increase significantly
as resolution is extended below our current limit.

As suggested by \citet{Bergstrom_et_al_99} and
\citet{Calcaneo_Moore_00}, if these subhaloes are sufficiently
concentrated, they can make a substantial contribution to the total
annihilation flux from the halo. Just as for the main halo, the
critical question regards the structure in their inner regions -- in
particular, whether they contain more or less mass at the highest
densities than does the core of the main halo. Recently
\citet{Hayashi_et_al_03} carried out high-resolution simulations of the
tidal stripping of satellites to assess how their internal structure
is affected by the removal of the outer material. Their results show
clearly that the stripping process reduces the density of an accreted
object at {\it all} radii, not just in its outer regions. Thus tidal
effects progressively lower the annihilation luminosity of an accreted
system. If its $\gamma$-ray flux was convergent in the inner regions
while it was an independent system, then it converges even more
rapidly once it has become a partially stripped `satellite'. Both
the individual satellite simulations of \citet{Hayashi_et_al_03} and
the results plotted in SWTS, suggest that the inner structure of halo
substructures corresponds to density profiles {\it shallower} than NFW.

We study this point in more detail in the left-hand panel of Fig.~\ref{figure:subhalocurves} which shows the circular
velocity profiles for a set of representative subhaloes in GA3n. This
can be compared with Fig.~2 of SWTS which gives a similar plot for the
9 times lower resolution simulation GA2 except that we here take only gravitationally bound particles into account. Clearly, the parabolic fitting formula
suggested by SWTS provides an excellent characterisation of circular
velocity curves in this higher resolution case also. As before, the
circular velocity curves for the subhaloes have substantially narrower
peaks than the curve for the main halo.  Indeed, the values of the
width parameter $a$ for GA3n subhaloes are very similar to those for
GA2 subhaloes. The shape convergence for subhalo density profiles is
demonstrated in the right-hand panel of Fig.~\ref{figure:subhalocurves}. We identify corresponding
subhaloes (i.e. subhaloes which formed from the infall of the same
progenitor object onto the main halo) in GA2n and GA3n. Then we plot a
$(a, r_{max})$ point for each of the two simulations and join the
symbols by a line. The agreement of the values found is quite good
and there is no systematic trend in either parameter as the mass
resolution is increased by an order of magnitude. The results of
\citet{Hayashi_et_al_03} confirm that subhalo centres are significantly
{\it less} cuspy either than the objects from which they formed or
than isolated haloes of similar mass. Their subhalo circular velocity
curves are very well fit by our parabolic formula and require similar
values of $a$. This agreement for two different simulation techniques
and over a resolution range of an order of magnitude suggests that the
reduced concentration of simulated satellites is unlikely to be an
effect of numerical relaxation, but more likely reflects the influence
of tidal shocking on the inner regions of satellite subhaloes.

We now proceed to estimate the annihilation luminosity of our
simulations directly in order to evaluate how much of the flux is
contributed by the various different parts of the system. To do this
we evaluate the `astrophysical' part of equation
(\ref{equation:flux}) in the form
\begin{equation}
J = \int_V \rho_{DM}^2 \, dV = \sum_{i=1}^{N_{200}} \rho_i \, m_i\, , 
\label{equation:astrophysicalpart}
\end{equation}
where $\rho_i$ is an estimate of the DM density at the position of the
$i$th particle, $m_i$ is its mass, and $N_{200}$ is the total number
of particles within $r_{200}$. With this representation of the flux it
is easy to evaluate the contribution from any sub-element of the halo
simply by restricting the sum to the relevant particles. We will
consider how these $J$s should be converted into $\gamma$-ray
detectability limits for various WIMP parameters in Section 5.

The quality of our estimate of $J$ will obviously depend strongly on
the quality of our estimates of DM density. For a given simulation we
would like these estimates to have the maximum possible resolution. We
have elected to determine the $\rho_i$ by Voronoi-tessellation. This
procedure uniquely divides three dimensional space into convex
polyhedral cells, one centred on each particle. A cell is defined to
contain all points closer to its particle than to any other. The
density estimate for each particle is then its mass divided by the
volume of its cell. We have used the publicly available package {\sc qhull}\footnote{www.geom.umn.edu/software/qhull/} to make these
estimates. One major advantage of this scheme in comparison, say, to
density estimation with an SPH kernel is that it is parameter-free and
has very high resolution; the density estimate for each particle is
determined by the position of its few nearest neighbours. Another is
that it is unbiased and that the sum $\sum_i^{N_{200}} m_i/\rho_i$
recovers the full volume.

\begin{figure}
\centering
\includegraphics[width=84mm,height=84mm]{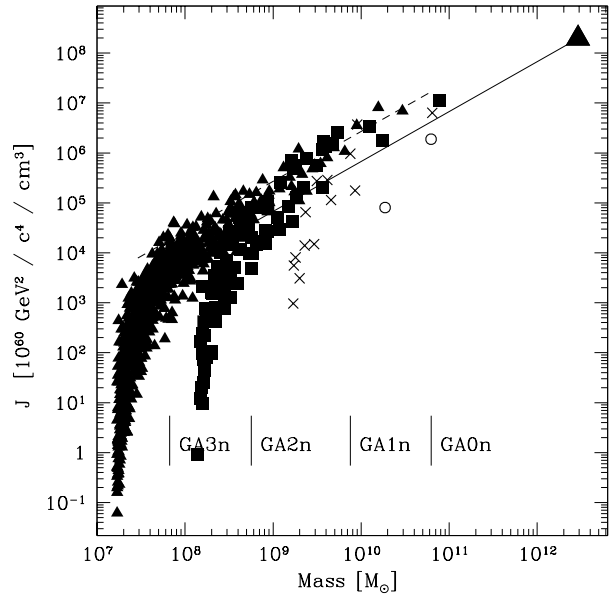}
\caption{The sum $J$, which is proportional to the expected
annihilation luminosity, is plotted as a function of subhalo mass for
all subhaloes with more than 50 particles in GA0n (circles), GA1n
(crosses), GA2n (dots) and GA3n (triangles). The value of $J$ for the
main halo as a whole is indicated by the larger triangle. Subhaloes
with the same $J/M$ as the GA3n main halo would lie on the solid
line. The dashed line corresponds to a 4 times larger value of
$J/M$. Vertical lines indicate subhalo masses corresponding to 200
particles in each of the four simulations. Above these limits
subhaloes in all four simulations have similar values of $J/M$ with no
obvious trend with subhalo mass. \label{figure:subhaloluminosity}}
\end{figure}

A direct numerical evaluation of $J$ using equation
\ref{equation:astrophysicalpart} will differ from the value obtained
by carrying out the appropriate integral over the circular velocity
curves of Fig.~\ref{figure:rotationcurves}. This is because any
deviation from strict spherical symmetry results in the sum of
$\rho_i m_i$ over a spherical shell being larger than the product of
its total mass and its mean density. Thus flux `enhancements' will
result from bound subhaloes, from unbound streams, from the overall
flattening of the halo and from noise in the density estimates due to
discreteness effects. The last contribution is easily estimated. In
any region where the mean particle separation is small compared to the
length-scale for density variations, our Voronoi scheme will give 22 per cent
more flux than would be obtained using the average density. For an
ellipsoidal NFW halo with axial ratios 1:1.2:1.8, similar to what we
measure in the inner regions of our simulated haloes, the enhancement due
to the flattening is about 15 per cent. The enhancement due to bound
structures is estimated explicitly below.

If subhaloes were simply scaled down copies of the main halo with
$r\propto M^{1/3}$, then their fractional contribution to the
annihilation luminosity would be the same as their fractional
contribution to the mass, i.e. roughly 5 per cent. However, the algorithm
{\sc subfind} bounds substructures at the point where their density
equals the local density of material within the main halo. As a
result, the internal structure of a subhalo cannot be similar to the main
halo as a whole, but might be similar to that part of it which lies
interior to the position of the subhalo (scaled down in size by the cube
root of the ratio of the substructure's mass to the total halo mass
within a sphere passing through it). If such self-similarity were
actually to hold then the annihilation luminosity per unit mass
(i.e. the quantity $J/M$) would be the same for the subhalo as for the
main halo interior to its position. In fact, however, the study of
\citet{Hayashi_et_al_03} shows that the density of a satellite at radii
approaching its tidal limit is reduced by a substantially greater
factor than that in the regions near the peak of its circular velocity
curve. This effect increases the luminosity per unit mass of a
substructure relative to the main halo interior to its
position. Higher values might also be expected because subhaloes had
lower mass progenitors at high redshift than did the main halo, and so
began life with higher concentration (see
\citet{Navarro_Frenk_White_97} and \citet{Bullock_et_al_01} for
estimates). On the other hand, we have argued above that tidal effects
also reduce the concentration of the inner core of subhaloes, thus
reducing their annihilation luminosities. The $J/M$ values for
substructures reflect the combination of all these effects.

Fig.~\ref{figure:subhaloluminosity} shows the sums $J$ -- which are
proportional to annihilation luminosity -- as a function of subhalo
mass for all four of our simulations and for subhaloes containing at
least 50 particles. A point is also plotted for the main halo as a
whole. If subhaloes were similar to the main halo, they would lie
along the solid line of constant $J/M$. Clearly, $J/M$ is larger for
the subhaloes. Vertical lines in the plot indicate subhalo masses
corresponding to 200 particles for each of the four simulations.
Above these limits the values of $J$ for subhaloes in GA0n, GA1n and
GA2n appear to converge approximately to those found in GA3n. This
suggests that the GA3n results may also be reliable for subhaloes with
more than 200 particles. Such GA3n subhaloes have $J/M$ values
typically 4 times larger than the main halo as a whole and twice as
large as the part of the main halo interior to their position. The
other effects discussed above presumably account for the remaining
factor. Note that there is no indication that $J/M$ depends on subhalo
mass for subhaloes with more than 200 particles. This implies that the
total luminosity from subhaloes, like the total subhalo mass, is
dominated by the largest objects.

\begin{figure*}
\centering
\includegraphics[width=84mm]{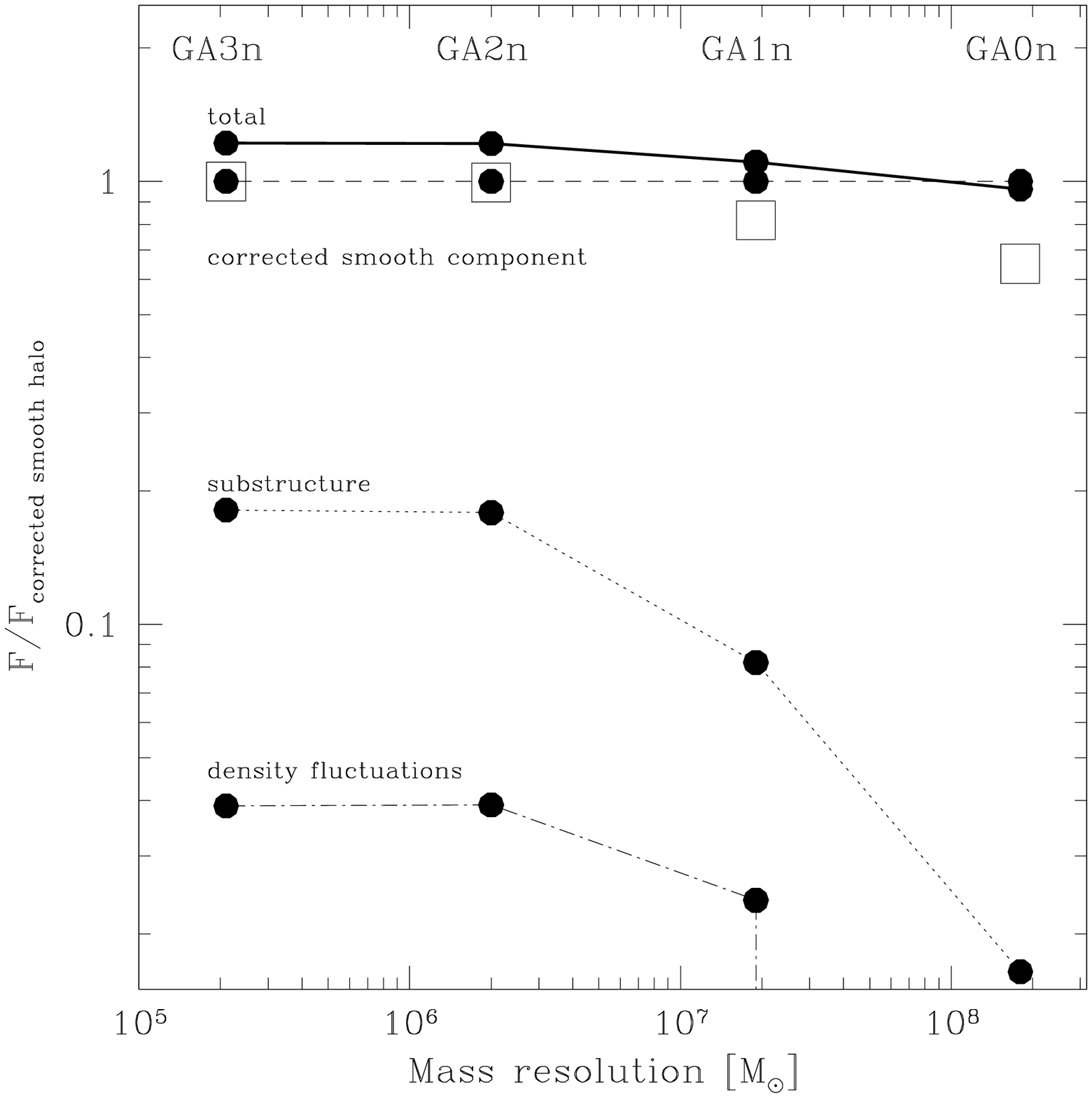}
\includegraphics[width=84mm]{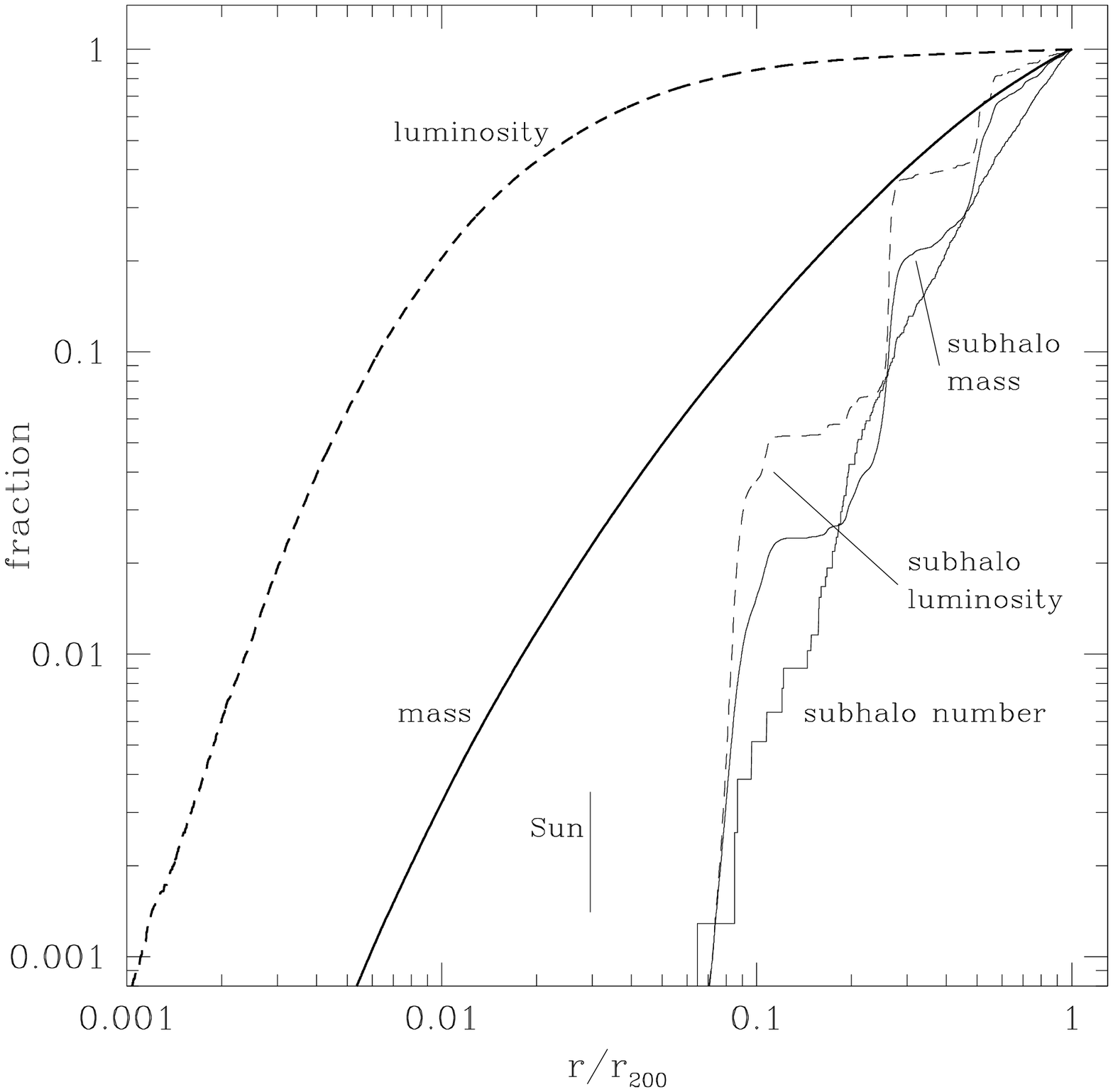}
\caption{{\it Left-hand panel}: The luminosities of different halo
components as a function of mass resolution. The values for each
simulation are scaled so that the luminosity of its smooth halo is
unity after correction for Poisson discreteness and flattening. The
large squares show the luminosities of these smooth halo components in
units of the value found in GA3n. {\it Right-hand panel}: Cumulative
luminosity (dashed) and mass (solid) for the GA3n main halo (thick)
and for the subhaloes with more than 50 particles (thin) as a function
of radius. \label{figure:luminosity}}
\end{figure*}

The left-hand panel of Fig.~\ref{figure:luminosity} shows the contributions
of different components to the total estimated annihilation luminosity
of our simulated haloes. The smooth halo component is shown by a
dashed line. This was calculated from the circular velocity curves of
Fig.~\ref{figure:rotationcurves} and was corrected up by 15 per cent to
account for the fact that the main halo is ellipsoidal rather than
spherical, and by 22 per cent to account for discreteness noise in our
Voronoi density estimates. The total luminosity from bound subhaloes
is indicated by a dotted line. The remainder of the total halo
luminosity is then assumed to come from unbound substructure and is
indicated by a dot-dashed line. All values are given in units of the
corresponding smooth halo luminosities. The values of the latter,
relative to GA3n, are indicated by boxes in the figure. The close
agreement of the circular velocity curves for GA3n and GA2n results in
identical predictions for the smooth halo luminosity. The values
predicted for GA1n and GA0n are only smaller by 20 and 35 per cent
respectively, suggesting that convergence is approximately achieved in
the inner regions even for relatively low mass resolution. The reason
is simply that the simulations predict the half-light radii of haloes
to be relatively large (8.6 kpc in GA3n) in comparison with the
nominal resolution.

In GA3n, the total luminosity is a factor of 1.7 -- the `clumpiness
factor' -- times the value for a smooth spherical halo with the same
circular velocity curve. Of this 70 per cent increase, 25 per cent is due to bound
substructures with 10 or more particles, 22 per cent to Poisson discreteness
noise, 15 per cent to the flattening of equidensity surfaces in the inner
halo and 8 per cent to unbound fluctuations. Although the fraction of the
annihilation luminosity contributed by substructure is very similar in
GA2n and GA3n, it is clear that this does not indicate absolute
convergence.  The total mass in substructures in GA2n is 1.2 times
larger than in GA3n (a consequence of the chance inclusion of a couple
of substructures in GA2n which lie just outside $r_{200}$ in GA3n). On
the other hand, the luminosity per unit mass of substructures is a
factor of 1.11 larger in GA3n than in GA2n. In addition, the
luminosity of the inner 10 per cent of $r_{200}$ (which dominates the
luminosity of the smooth halo) is 5 per cent larger in GA3n than in GA2n.

In order to get an idea of an upper limit of the additional luminosity
which might be found at higher resolution it is instructive to
extrapolate the variation between GA1n and GA2n down to a mass
resolution of one solar mass. (Note that GA3n lies well {\it below}
this extrapolation.)  Even at such high resolution, the total
luminosity is predicted to be only about 3.0 times that of the smooth
halo in GA3n.

\citet{Lake_90} and \citet{Bergstrom_et_al_99} suggested that if a
dark matter substructure happens to be close to the observer, it might
be more easily detected than the Galactic Centre itself. This
possibility was judged plausible by \citeauthor{Tasitsiomi_Olinto_02}
(\citeyear{Tasitsiomi_Olinto_02}; hereafter TO) who assumed subhaloes
to be distributed through the Galactic halo like the DM itself and
tried various models for their internal structure. For the internal
structure predicted by our simulations, however, it is very unlikely
that any substructure will outshine the Galactic Centre. The most
massive and most luminous substructures are rare and tend to avoid the
inner Galaxy.  They presumably correspond to the known satellites of
the Milky Way (see SWTS), the nearest of which is Sagittarius, 24 kpc
from the Sun. The greater abundance predicted for less massive
substructures is insufficient to compensate for their lower predicted
luminosities -- the chance that the received flux is dominated by an
unexpectedly nearby low-mass substructure is predicted to be very low.

These issues are clarified in the right-hand panel of
Fig.~\ref{figure:luminosity}, where we plot cumulative radial
distributions for total mass and total annihilation luminosity {\it
exclusive of substructure}, as well as for the number, mass and
annihilation luminosity of subhaloes. While the diffuse luminosity is
much more strongly concentrated towards the Galactic Centre than the
mass, all properties of the substructure are more {\it weakly}
concentrated. In addition, since $J/M$ is independent of subhalo mass,
which in turn is almost independent of distance from the Galactic
Centre, the distributions of substructure number, mass and luminosity
are all rather similar. The latter two are much noisier than the first
because most of the mass and most of the luminosity come from the few
most massive subhaloes. At 8 kpc, the Sun's galactocentric radius lies
in the region where most of the diffuse annihilation radiation
originates, but well inside any of the resolved subhaloes in GA3n (the
first is at $R=17.2$~kpc) and even further inside any of the more
massive subhaloes (the first is at $R=70$~kpc).

Whereas Fig.~\ref{figure:luminosity} was constructed directly from GA3n,
based on our Voronoi estimate of $J/M$ for each particle, we obtain
almost identical results if we instead use our SWTS model fits to main
halo and subhalo circular velocity curves and assume that $J/M$ is 4
times the value for the main halo for all subhaloes which are too
small for circular velocity curves to be estimated. Again, this
suggests that GA3n has high enough resolution to get reliable results
for the problem at hand.

\section{Detectability}
\label{section:detectability}

\begin{figure*}
\includegraphics[width=84mm]{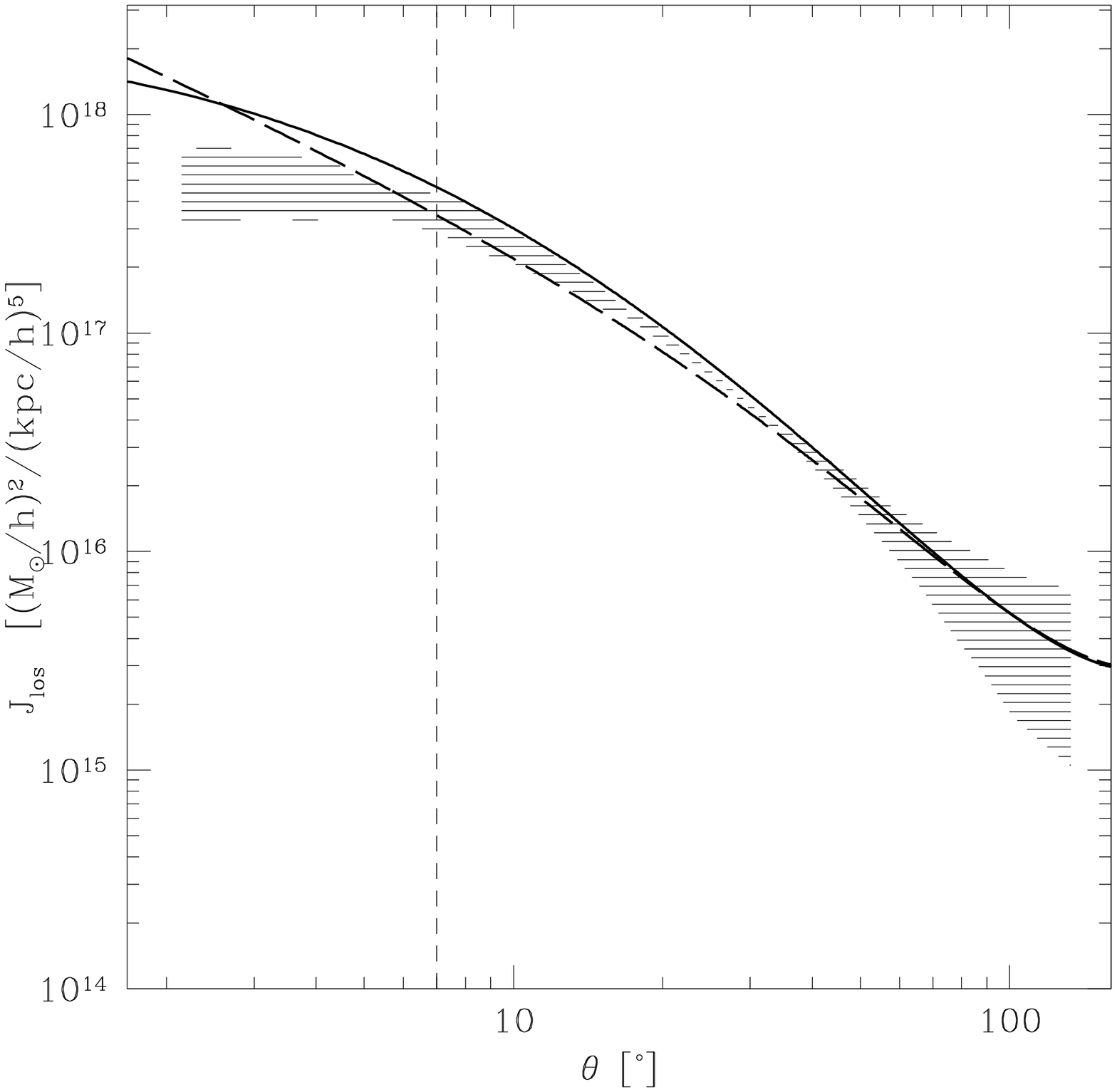} 
\includegraphics[width=84mm]{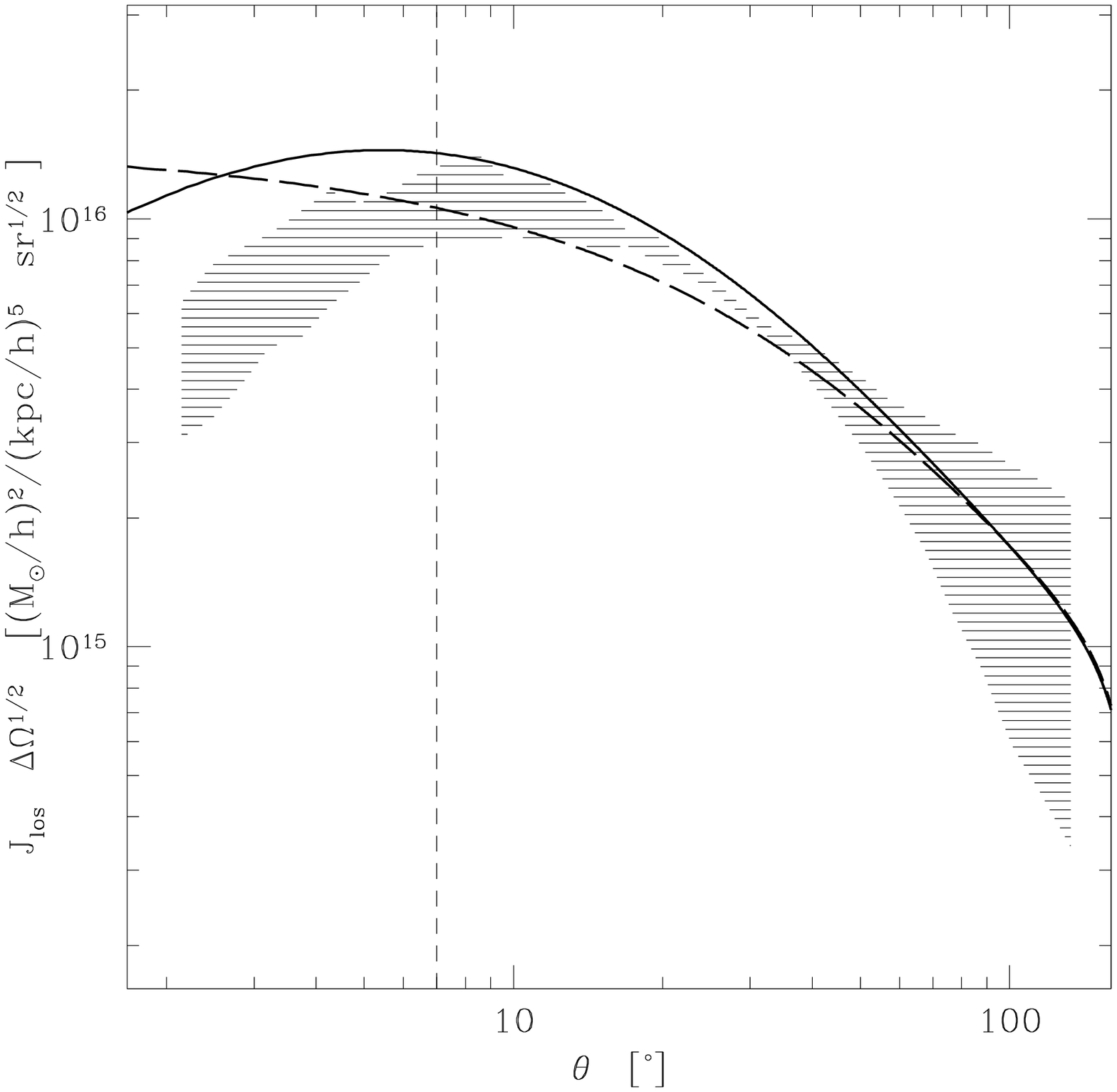} 
\caption{Mean predicted surface brightness of
annihilation radiation as a function of angular distance from the
Galactic Centre. The hatched regions enclose profiles estimated
directly from six different `solar' positions 8kpc from the centre
of GA3n, while the two curves gives results based on SWTS (solid) and
NFW (dashed) fits to the circular velocity curve, together with an
enhancement factor of 1.7. In the left-hand panel the surface
brightness profiles are plotted directly, whereas on the right they
are multiplied by one power of the angle $\theta$ to represent the
signal-to-noise expected in an observation of fixed duration of a
region whose size increases in proportion to $\theta$, (specifically [$\theta$,$1.01 \theta$]). Vertical dashed
lines indicate the angle subtended by the gravitational softening
length at the distance of the Galactic Centre.
\label{figure:lineofsight}}
\end{figure*}

For our highest resolution simulation GA3n we can make artificial sky
maps of the annihilation radiation by choosing appropriate positions
for the Sun within the model. Fig.~\ref{figure:lineofsight} shows the
result of this exercise based on six possible solar positions. Even
though we average the predicted surface brightness around circles of
fixed Galactocentric angle, there is significant variation among the
resulting profiles. This is primarily a consequence of the prolate
structure of the inner regions, clearly visible in Fig.~1. The
profiles flatten out within about 10$^\circ$, quite possibly as a
consequence of poor numerical resolution. Prima facie this seems
plausible since the angular scale corresponding to our softening
length (the vertical lines in the plots) is only 4 or 5 times smaller
than the radius where the profiles bend. Some indication of the
strength of this effect is given by the two curves. These indicate
predictions based on SWTS (solid) and NFW (dashed) fits to the
circular velocity profile of GA3n, corrected up by a clumpiness factor
of 1.7. Both inward extrapolations predict substantially more flux
within a few degrees than the direct numerical estimates. 

It is important to note, however, that the area potentially available
for a measurement at distance $\theta$ from the Galactic Centre
increases as $\theta$ (for $\theta \ll \pi/2$). As a result, the counts available to detect a
signal vary as $\theta^2$ times the profiles shown in the left panel
of Fig.~\ref{figure:lineofsight} while, for a uniform background, the
noise against which the signal must be detected grows only as
$\theta$. Thus the potential signal-to-noise for a detection, shown in the
right-hand panel, is given by $\theta$ times the profile. This
function is quite flat out to 20$^\circ$, both for the directly
measured profiles and for our two alternative fitting formulae. This
has two consequences: (i) since for many observations the background
is higher in the direction of the Galactic Centre, it may be
advantageous to observe at large $\theta$ if one has a detector with
sufficient field of view; and (ii) the estimates of detectability which
we give below for detectors with a wide field of view are not greatly
affected by the resolution of our simulation.

\begin{figure}
\includegraphics[width=84mm]{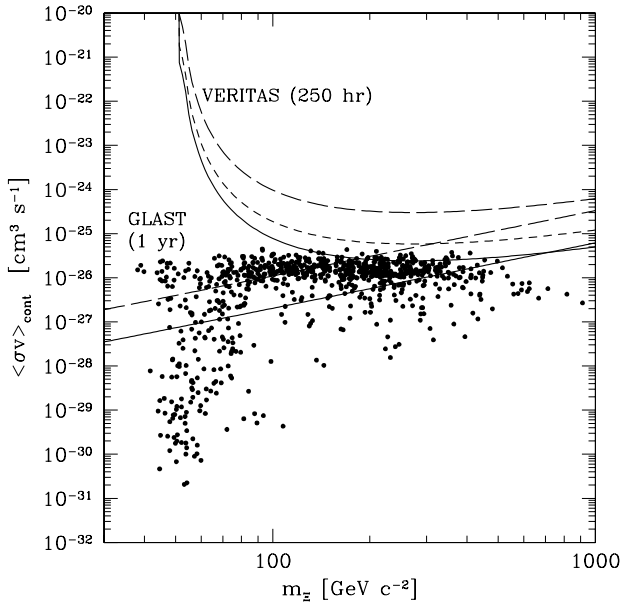} 
\caption{MSSM models of cosmological interest (dots) and 3-$\sigma$
detection limits for VERITAS and GLAST. For VERITAS
the limits are shown for a pointing at the centre of the Milky Way,
assuming an NFW profile (solid) and an SWTS profile (short
dashes). The lower solid line gives estimated limits for GLAST for
a larger area observation of the inner Galaxy which avoids regions of
high contamination by diffuse Galactic emission. Limits for a pointing
at the brightest high latitude subhalo are shown for both telescopes
using long dashes. The brightest subhalo was chosen from the 6
artificial skies used in making Fig.~\ref{figure:lineofsight}.
\label{figure:detectability}}
\end{figure}

Using these results, we can check if the inner regions of the Milky
Way or a substructure halo close to the Sun might be detectable with
next generation $\gamma$-ray telescopes. We use the excellent package
{\sc darksusy}\footnote{www.physto.se/$\sim$edjso/darksusy/} to
compute the cross-sections $\left<\sigma v\right>$ and neutralino
masses $m_{\chi}$ for a Monte Carlo sampling of the MSSM parameter
space. The results are shown in Fig.~\ref{figure:detectability}. From
roughly two million models randomly picked out of the parameter space,
19~421 did not violate current accelerator bounds. Of these, 825 result
in relic densities of cosmological interest,
i.e. $0.17<\Omega_{DM}<0.43$, the 95 per cent confidence interval quoted by
\citet{Spergel_et_al_03}. For sampling the MSSM parameter space we followed the
choices of TO. For given telescope and observation parameters -- i.e. the
effective area $A_{eff}$, the integration time $t$, the angular
resolution $\sigma_{\theta}$, the radius of the field of view
$\theta_{max}$, the effective background count rate and the
significance required for detection $M_s$ -- the smallest detectable
cross-section $\left<\sigma v\right>$ can be computed as a function of
the neutralino mass $m_{\chi}$ \citep[ TO]{Bergstrom_et_al_98,
Baltz_et_al_00}. In computing the expected signal, the volume of
integration in equation (1) has to be taken as the volume of the halo
contained within the chosen detection cell which will be the
resolution element of the telescope for highly concentrated sources
but may be much larger for diffuse sources such as the emission
predicted in Fig.~\ref{figure:lineofsight}. We concentrate on
estimating the $\gamma$-ray continuum signal which is easier to detect
than the line signal \citep[TO]{Baltz_et_al_00}.

Averaged over a gaussian beam of width
$\sigma_{\theta}$=0.1$^{\circ}$, the resolution element for the
telescopes listed in Table~\ref{table:telescopes}, the line-of-sight integral of the
square of the mass density (equation \ref{equation:averagedlineofsight}) in the direction of the Galactic Centre
takes values 5.2 $\times 10^{25}$ and 1.8 $\times 10^{24}$
GeV$^2$~$c^{-4}$~cm$^{-5}$ for inward extrapolations of our NFW and SWTS
fits to the main halo circular velocity curve. The large difference
reflects the fact that this estimate is sensitive to density values
far inside the region resolved by our simulations.
Fig.~\ref{figure:rotationcurves} suggests that the lower value
obtained from the SWTS extrapolation is more likely correct. To
estimate the maximum plausible brightness for a subhalo, we examined
the six artificial skies used to make Fig.~\ref{figure:lineofsight}
and identified the brightest subhalo in each after convolution with a
0.1$^{\circ}$ beam. The beam-averaged line-of-sight integral of
density squared for the (apparently) brightest substructure in these
six realisations is $4.9\times 10^{23}$ GeV$^2$~$c^{-4}$~cm$^{-5}$.

To estimate detectability, we have to specify the appropriate
background levels. We account for the electron-induced background in
ACT observations, and for the extragalactic background emission in
space-based observations. We neglect any hadronic background.  All
observations at low latitudes and in the general direction of the
Galactic Centre are in addition affected by the diffuse Galactic
$\gamma$-ray emission. Although this contribution can be neglected for
ACT observations, it is the dominant background in these directions
for space experiments like GLAST. In combination with the
results of Fig.~\ref{figure:lineofsight}, this implies that the best
signal-to-noise is expected for an observation of a broad broken annulus which
{\it excludes} the Galactic Center and the Galactic Plane. If we
assume that the diffuse galactic component drops to the level of the
extra-galactic background beyond 30$^\circ$ from the centre and 10$^\circ$ 
from the plane, a signal-to-noise ratio can be achieved which
is about 12 times better than that for a 0.1$^\circ$ beam in the direction of
the Galactic Centre (for an assumed NFW profile). This result is
spectacular: the density profile of the DM halo in these regions is
well established from the simulations and the prediction becomes {\it
independent} of numerical uncertainties in the innermost structure of
CDM haloes.

These results are shown in Fig.~\ref{figure:detectability}. The solid
curve gives our estimated lower limit on the cross-section for a
3-$\sigma$ detection of the Galactic Centre in a 0.1$^\circ$ beam for a
250-h observation with the planned ACT VERITAS. This particular
calculation extrapolates to small radii using the NFW model of
Fig.~\ref{figure:lineofsight} and the signal-to-noise is then maximised for the
smallest resolved detection cell. The short-dashed curve is the
corresponding limit for inward extrapolation using the SWTS model. In
this case the signal-to-noise is maximised using a detection cell of radius 1.75$^\circ$, 
corresponding to the full field of view of VERITAS. The
corresponding lower limit for detection of the brightest substructure
in our six artificial skies, again for a 1.75$^\circ$ beam, is shown by 
the long dashed curve. We
assume the inner density structure of this object to be correctly
described by our SWTS model fit. A 250-h observation may just be
enough detect the Galactic Centre, at least for a few of the plausible
MSSM models. Detection of substructure appears out of reach unless our
simulations have grossly underestimated the central concentration of
subhaloes.

Fig.~\ref{figure:detectability} also shows cross-section limits for a
1-yr exposure with the satellite telescope GLAST. The straight
long-dashed line is the limit for detecting the brightest satellite,
assuming this to be outside the region with strong diffuse Galactic
emission and using a detection cell of radius 5$^\circ$ corresponding to the 
the peak signal-to-noise angle. The straight solid line gives the limit for detecting
annihilation radiation in an annulus between 25$^\circ$ and 35$^\circ$ from
the Galactic Centre but excluding the region within 10$^\circ$ of the
Galactic Plane. We assume that the total diffuse Galactic emission in this region
is zero. The results here are quite encouraging. The inner Galaxy should be detectable for
most allowable MSSM parameters, while the brightest substructure is
also detectable for many of them (for TO's implicitly adopted prior on 
the MSSM parameter space).

Whereas the field of view of GLAST covers almost a fifth of the
full sky, the smaller field of view of VERITAS allows
observation of only one object at a time. In addition, ACTs can only
operate about 6 h per night. For these reasons we consider
exposure times of 1 yr for GLAST and 250 h for VERITAS to be large 
but feasible.  

\begin{table}
\caption{Simplified telescope specifications \label{table:telescopes}}
\begin{center}
\begin{tabular}{l l c c c c }
   \hline
    & $A_{eff}$  & $E_{th}$ & $\sigma_{\theta}$ & $\theta_{max}$\\
    & [m$^2$] & [GeV $c^{-2}$] & [$^\circ$] & [$^\circ$]\\
    \hline
    VERITAS  & 10$^4$ & 50 & 0.1 & 1.75\\
    GLAST    & 0.8 & 0.02 & 0.1 & 53\\
    \hline
\end{tabular}
\end{center}
\end{table}

\section{Discussion}
\label{section:Discussion}
We have directly estimated the $\gamma$-ray emissivity of the halo of the Milky
Way using high resolution simulations of its formation in a
standard $\Lambda$CDM universe. A series of resimulations of the same
DM halo at different mass resolution allows us to check explicitly for
numerical convergence in our results. We find that the resolution
limit of our largest simulation is almost an order of magnitude
smaller than the half-light radius for the annihilation radiation, and
that our estimates of the total flux are almost converged. We
argue that the annihilation radiation from substructure within the
Galactic halo is dominated by the most massive subhaloes, is
concentrated in the outer halo, and is less in total than the
radiation from the smooth inner halo. For the most massive subhaloes
our convergence study indicates sufficient resolution in our
best simulations to get robust estimates of their internal structure.
An important result is that subhalo cores are less concentrated
both than that of the main halo and than those of their progenitor
haloes. This confirms earlier results by SWTS and Hayashi et al
(2003) and apparently reflects the influence of tidal shocking on
subhalo structure.

We find that 15 per cent of the total flux in our highest-resolution
simulation is coming from gravitationally bound subhaloes and that no
more than about 5 per cent can be assigned to other small-scale density
fluctuations. Some of our results do rely strongly on the density
behaviour we infer from our simulations for the innermost regions both
of the main halo and of the subhaloes. This subject is still
controversial, although the most detailed convergence study to date
agrees quite well with our results for the centre of the main halo
\citep{Power_et_al_03}. As already noted, our subhalo structure agrees
well with that found by Hayashi et al (2003). The disagreement between
our results and other theoretical work on annihilation luminosities
(Calcaneo-Rodan \& Moore 2000, Tasitsiomi \& Olinto 2002, Taylor \&
Silk 2003) can be traced to the fact that the density profiles adopted
in these papers are incompatible with those we measure in our
simulations, particularly for subhaloes. It may be relevant that
observational data on dwarf galaxies also speak in favour of dark
matter density profiles with low concentrations or cores
\citep{DeBlok_et_al_01} although again the situation is
controversial here.

To estimate the fluxes expected for deep integrations with upcoming experiments, it is
necessary to extrapolate density profiles down to scales at least an
order of magnitude below those where they can be reliably estimated in
our simulations. Clearly, this introduces substantial uncertainty. Our
results suggest that the central regions of the Galaxy will be
intrinsically more luminous than the brightest substructure by a
factor of at least 10, and apparently more luminous by a factor
approaching one thousand. The angular scale associated with the
central emission will be several tens of degrees while that associated
with the substructure will be a few degrees.  This suggests that it
may be worthwhile to investigate detection strategies which are
sensitive to large-scale diffuse emission. Notice that since our
results imply that the most apparently luminous subhaloes will also be
among the most massive, it is likely that the brightest substructure
source will be identified with one of the known satellites of the
Milky Way. The closest of these is the Sagittarius dwarf at a distance
of 24 kpc, but it may well be outshone by the LMC at a distance of 45
kpc. Both are sufficiently far that they will be much fainter (and
smaller in angular size) than the main halo source which is centred
only 8 kpc away.

Following TO, and using {\sc darksusy}, we checked for detectability of
the inner Milky Way with VERITAS and GLAST, examples of
ground- and space-based next-generation telescopes, respectively. If we
extrapolate our simulated density profiles inwards using an NFW fit,
VERITAS can probe into the parameter ranges in which a minimal
supersymmetric extension of the Standard Model could provide a Dark
Matter candidate with the observed cosmic density.
Unfortunately, extrapolating inwards using our SWTS
fit, which appears to provide a better description of our simulations,
results in lower predicted fluxes, undetectable for VERITAS.

By searching for extended emission outside the central region where
diffuse Galactic $\gamma$-ray emission is dominant GLAST
can probe a large region of possible MSSM models. This result is based
on the DM distribution in regions where the simulations have reliably
converged, and so should be robust. It is {\it independent} of the
exact structure of the DM in the innermost regions.

Our simulations suggest that the flux from the inner Galaxy will
outshine the brightest substructure by a large factor. Nevertheless,
for certain MSSM models some of the most massive substructure haloes 
might be detectable with GLAST. Clearly the most massive and
nearest {\it known} satellite galaxies are the primary targets for
observation.

\section*{Acknowledgments}

We thank Eric Nuss, Karsten Jedamzik, Argyro Tasitsiomi, Pasquale Blasi, Paolo Gondolo and Torsten En{\ss}lin for many enlightening discussions.

\appendix
\section[]{Detectability Computation}
We summarise briefly how the detectability lines of
Fig.~\ref{figure:detectability} were computed following
\citet{Bergstrom_et_al_98}, \citet{Baltz_et_al_00},
\citet{Ullio_et_al_02} and TO. The solid angle corresponding to the
angular resolution $\sigma_{\theta}$ of the telescope may be written as:
\begin{equation}
  \Delta\Omega = 2 \pi \left[ 1- \cos(\sigma_{\theta})\right],
\end{equation}
while the number of continuum photons arriving on the telescope is
\begin{equation}
  N_{annihilation} =  A_{eff} \ t \frac{\ N_{cont}}{2} \ \frac{\ \left<\sigma v\right>}{\ m_{\chi}^2} \ \frac{\Delta\Omega}{4\pi} G_{los,\Delta\Omega}.
\end{equation}
Here, $G_{los,\Delta\Omega}$ stands for the line-of-sight integral of
the DM distribution averaged over the solid angle $\Delta\Omega$:
\begin{equation}
G_{los,\Delta\Omega} = \frac{1}{\Delta\Omega} \ \int_{\Delta\Omega} {\mbox d}\Omega \ \ \int_{los} \rho_{DM}^2(l) \ {\mbox d}l .
\label{equation:averagedlineofsight}
\end{equation}
Values for $G_{los,\Delta\Omega}$ are given in Section
\ref{section:detectability}. TO give an approximate formula to compute
the number continuum photons from one annihilation:
\begin{equation}
N_{cont}(E_{\gamma}>E_{th}) = \frac{5}{6} x^{3/2} - \frac{10}{3} x + 5 x^{1/2} + \frac{5}{6} x^{-1/2} - \frac{10}{3}
\end{equation}
Here $x =E_{th}/m_{\chi}$ is the quotient of the energy threshold of
the telescope and the neutralino mass. Finally, the significance of a
detection $M_s$ is given by the number of detected photons from DM
annihilations over the square root of the background:
\begin{equation}
M_s \leq \frac{N_{annihilation}}{\sqrt{N_{background}}}.
\end{equation}
This allows us to compute the minimal detectable cross-section
$\left<\sigma v\right>_{min}$ as a function of the mass of the
neutralino $m_{\chi}$ via:
\begin{equation}
   \left<\sigma v\right>_{min} = \frac{2 \ M_s \ m_{\chi}^2 \  \sqrt{N_{background}}}{N_{cont} \ A_{eff} \ t \  \frac{\Delta\Omega}{4\pi} \ G_{los,\Delta\Omega}}
\end{equation}
The detectability scales with $M_s$, $t^{-1/2}$ and $A_{eff}^{-1/2}$. The background counts (hadronic, electron-induced, diffuse-galactic (for the centre of the galaxy) and extra-galactic) are taken from \citet{Bergstrom_et_al_98} and \citet{Baltz_et_al_00}:
\begin{equation}
\frac{{\mbox d}N_{h}}{{\mbox d}t \ {\mbox d}A\ {\mbox d}\Omega} = 6.1 \times 10^{-3}  \left(\frac{E_{th}}{1\ {{\mbox {GeV/c}}^2}} \right)^{-1.7} {\mbox  {cm}}^{-2} \ {\mbox s}^{-1}\ {\mbox sr}^{-1}
\end{equation}
\begin{equation}
\frac{{\mbox d}N_{e}}{{\mbox d}t \ {\mbox d}A\ {\mbox d}\Omega} = 3.0 \times 10^{-2}  \left(\frac{E_{th}}{1 \ {{\mbox {GeV/c}}^2}} \right)^{-2.3} {\mbox  {cm}}^{-2} \ {\mbox s}^{-1}\ {\mbox sr}^{-1}
\end{equation}
\begin{equation}
\frac{{\mbox d}N_{d}}{{\mbox d}t \ {\mbox d}A \ {\mbox d}\Omega} =5.1 \times 10^{-5}  \left(\frac{E_{th}}{1 \ {{\mbox {GeV/c}}^2}} \right)^{-1.7} {\mbox  {cm}}^{-2} \ {\mbox s}^{-1}\ {\mbox sr}^{-1}
\end{equation}
\begin{equation}
\frac{{\mbox d}N_{eg}}{{\mbox d}t \ {\mbox d}A \ {\mbox d}\Omega} = 1.2 \times 10^{-6}  \left(\frac{E_{th}}{1 \ {{\mbox {GeV/c}}^2}} \right)^{-1.1}{\mbox  {cm}}^{-2} \ {\mbox s}^{-1}\ {\mbox sr}^{-1}
\end{equation}

\begin{thebibliography}{29}
\expandafter\ifx\csname natexlab\endcsname\relax\def\natexlab#1{#1}\fi

\bibitem[{Baltz} et~al.(2000){Baltz}, {Briot}, {Salati}, {Taillet} \&
  {Silk}]{Baltz_et_al_00}
{Baltz} E.~A., {Briot} C., {Salati} C., {Taillet} R., {Silk} J., 2000, Physical
  Review, 61

\bibitem[{Bergstr{\" o}m} et~al.(1999){Bergstr{\" o}m}, {Edsj{\" o}}, {Gondolo}
  \& {Ullio}]{Bergstrom_et_al_99}
{Bergstr{\" o}m} L., {Edsj{\" o}} J., {Gondolo} P., {Ullio} P., 1999, Physical
  Review, 59

\bibitem[{Bergstr{\" o}m} et~al.(1998){Bergstr{\" o}m}, {Ullio} \&
  {Buckley}]{Bergstrom_et_al_98}
{Bergstr{\" o}m} L., {Ullio} P., {Buckley} J.~H., 1998, Astroparticle Physics,
  9, 137

\bibitem[{Bullock} et~al.(2001){Bullock}, {Kolatt}, {Sigad}
  et~al.]{Bullock_et_al_01}
{Bullock} J.~S., {Kolatt} T.~S., {Sigad} Y., et~al., 2001, \mnras, 321, 559

\bibitem[{Calcan{\' e}o-Rold{\' a}n} \& {Moore}(2000)]{Calcaneo_Moore_00}
{Calcan{\' e}o-Rold{\' a}n} C., {Moore} B., 2000, Physical Review, 62

\bibitem[{de Blok} et~al.(2001){de Blok}, {McGaugh} \&
  {Rubin}]{DeBlok_et_al_01}
{de Blok} W.~J.~G., {McGaugh} S.~S., {Rubin} V.~C., 2001, \aj, 122, 2396

\bibitem[{Dekel} et~al.(2002){Dekel}, {Arad}, {Devor} \&
  {Birnboim}]{Dekel_et_al_02}
{Dekel} A., {Arad} I., {Devor} J., {Birnboim} Y., 2002, in {
  astro-ph/0205448\/}

\bibitem[{Font} et~al.(2001){Font}, {Navarro}, {Stadel} \&
  {Quinn}]{Font_et_al_01}
{Font} A.~S., {Navarro} J.~F., {Stadel} J., {Quinn} T., 2001, \apjl, 563, L1

\bibitem[{Hayashi} et~al.(2003){Hayashi}, {Navarro}, {Taylor}, {Stadel} \&
  {Quinn}]{Hayashi_et_al_03}
{Hayashi} E., {Navarro} J.~F., {Taylor} J.~E., {Stadel} J., {Quinn} T., 2003,
  \apj, 584, 541

\bibitem[{Helmi} et~al.(2002){Helmi}, {White} \& {Springel}]{Helmi_et_al_02}
{Helmi} A., {White} S.~D., {Springel} V., 2002, \prd, 66, 63502

\bibitem[{Hernquist}(1990)]{Hernquist_90}
{Hernquist} L., 1990, \apj, 356, 359

\bibitem[{Klypin} et~al.(1999){Klypin}, {Kravtsov}, {Valenzuela} \&
  {Prada}]{Klypin_et_al_99}
{Klypin} A., {Kravtsov} A.~V., {Valenzuela} O., {Prada} F., 1999, \apj, 52, 82

\bibitem[{Lake}(1990)]{Lake_90}
{Lake} G., 1990, \nat, 346, 39+

\bibitem[{Metcalf} \& {Madau}(2001)]{Metcalf_Madau_01}
{Metcalf} R.~B., {Madau} P., 2001, \apj, 563, 9

\bibitem[{Moore} et~al.(1999){Moore}, {Ghigna}, {Governato}
  et~al.]{Moore_et_al_99}
{Moore} B., {Ghigna} S., {Governato} F., et~al., 1999, \apjl, 524, L19

\bibitem[{Navarro} et~al.(1997){Navarro}, {Frenk} \&
  {White}]{Navarro_Frenk_White_97}
{Navarro} J.~F., {Frenk} C.~S., {White} S. D.~M., 1997, \apj, 490, 493+

\bibitem[{Nusser} \& {Sheth}(1999)]{Nusser_Sheth_99}
{Nusser} A., {Sheth} R.~K., 1999, \mnras, 303, 685

\bibitem[{Peebles}(1980)]{Peebles_80}
{Peebles} P. J.~E., 1980, The Large-Scale Structure of the Universe, Princeton

\bibitem[{Power} et~al.(2003){Power}, {Navarro}, {Jenkins}
  et~al.]{Power_et_al_03}
{Power} C., {Navarro} J.~F., {Jenkins} A., et~al., 2003, \mnras, 338, 14

\bibitem[{Spergel} et~al.(2003){Spergel}, {Verde}, {Peiris}
  et~al.]{Spergel_et_al_03}
{Spergel} D.~N., {Verde} L., {Peiris} H.~V., et~al., 2003, ArXiv Astrophysics
  e-prints,  2209--+

\bibitem[Springel et~al.(2001{\natexlab{a}})Springel, White, Tormen \&
  Kauffmann]{Springel_et_al_01}
Springel V., White S. D.~M., Tormen G., Kauffmann G., 2001{\natexlab{a}},
  MNRAS, 328, 726

\bibitem[Springel et~al.(2001{\natexlab{b}})Springel, Yoshida \&
  White]{Springel_Yoshida_White_01}
Springel V., Yoshida N., White S. D.~M., 2001{\natexlab{b}}, New Astronomy, 6,
  79

\bibitem[{Stoehr} et~al.(2002){Stoehr}, {White}, {Tormen} \&
  {Springel}]{Stoehr_et_al_02}
{Stoehr} F., {White} S.~D.~M., {Tormen} G., {Springel} V., 2002, \mnras, 335,
  L84

\bibitem[{Subramanian} et~al.(2000){Subramanian}, {Cen} \&
  {Ostriker}]{Subramanian_et_al_00}
{Subramanian} K., {Cen} R., {Ostriker} J.~P., 2000, \apj, 538, 528

\bibitem[{Syer} \& {White}(1998)]{Syer_White_98}
{Syer} D., {White} S. D.~M., 1998, \mnras, 293, 337+

\bibitem[{Tasitsiomi} \& {Olinto}(2002)]{Tasitsiomi_Olinto_02}
{Tasitsiomi} A., {Olinto} A.~V., 2002, \prd, 66, 83006

\bibitem[{Taylor} \& {Silk}(2003)]{Taylor_Silk_03}
{Taylor} J.~E., {Silk} J., 2003, \mnras, 339, 505

\bibitem[{Tormen} et~al.(1997){Tormen}, {Bouchet} \&
  {White}]{Tormen_Bouchet_White_97}
{Tormen} G., {Bouchet} F.~R., {White} S. D.~M., 1997, \mnras, 286, 865

\bibitem[{Ullio} et~al.(2002){Ullio}, {Bergstr{\" o}m}, {Edsj{\" o}} \&
  {Lacey}]{Ullio_et_al_02}
{Ullio} P., {Bergstr{\" o}m} L., {Edsj{\" o}} J., {Lacey} C., 2002, \prd, 66,
  123502

\end{thebibliography}
\end{document}